\documentstyle[11pt,epsf,newpasp,twoside]{article}
\def\mpc {h^{-1} {\rm{Mpc}}}

\def\edcomment#1{\iffalse\marginpar{\raggedright\sl#1\/}\else\relax\fi}
\reversemarginpar

\begin{document}
\title{The Galaxy Environment of Quasars in the ${\bf z\simeq1.3}$ Clowes-Campusano Large Quasar Group}
 \author{C.P. Haines \& R.G. Clowes}
\affil{Centre for Astrophysics, University of Central Lancashire, \\Preston, PR1 2HE, UK}
\author{L.E. Campusano}
\affil{Observatorio Astr\'{o}nomico Cerro Cal\'{a}n, Departamento de Astronom\'{i}a, Universidad de Chile, Casilla 36-D, Santiago, Chile}

\begin{abstract}
The Clowes-Campusano Large Quasar Group (LQG) is the largest known structure at high redshifts, with at least 18 quasars at $z\simeq1.3$ forming a structure $\sim200\mpc$ across. We are conducting an ultra-deep optical and deep NIR study to examine the galaxy environment of quasars in this LQG to determine if, and how, LQGs trace large-scale structure at early epochs. 
We report significant associated clustering in the field of a $z=1.226$ quasar from this LQG in the form of a factor $\sim11$ overdensity of $I-K>3.75$ galaxies, and red sequences of \mbox{15--18} galaxies at $I-K\simeq4.3,V-K\simeq6.9$, indicative of a population of massive ellipticals at the quasar redshift.
 The quasar is located between two groups of these galaxies, with further clustering extending over \mbox{2--3 Mpc}. Within 30 arcsec of the quasar we find a concentration of blue \mbox{($V-I<1$)} galaxies in a band that bisects the two groups of red sequence galaxies. This band is presumed to correspond to a region of enhanced star-formation. We suggest that the merging of the two groups of red sequence galaxies has triggered both the quasar and the band of star-formation. Quasars at $z\simeq1.3$ are located in a variety of environments, but those with associated clustering are found on the cluster peripheries.
\end{abstract}

\section{Introduction}

Quasars have been used as efficient probes of high-redshift clustering because they are known to favour rich environments. Quasars may also trace large-scale structures at early epochs ($0.4\la z\la 2$) in the form of Large Quasar Groups (LQGs), which have sizes comparable to the largest structures at the present epoch. Several examples of LQGs are known (e.g. Webster 1982; Crampton, Cowley \& Hartwick 1989), of which the largest is the Clowes-Campusano LQG of 18 quasars at $z\simeq1.3$, with a maximal extent of $\sim200\mpc$ (Clowes \& Campusano 1991), making it the largest known structure at high redshifts. 

It has been suggested that LQGs trace large-scale structures at early epochs, and there is increasing evidence that this is the case, with associated Mg{\sc ii} absorbers, star-forming regions, and galaxy clustering all recently observed. However the relationship between the LQG quasars and mass (galaxies) appears to be rather complicated. Gas associated with the Clowes-Campusano LQG has been detected by Mg{\sc ii} absorption from observations of background quasars, with a $3.4\sigma$ overdensity at $1.2\leq z<1.3$ (11 observed, 4 expected). Optical and narrow-band (O{\sc ii}) imaging of quasars from the $z\simeq1.1$ Crampton et al. LQG find excesses of blue and emission-line galaxies around 10 of the 11 quasars, indicating that the quasars are located within regions of enhanced star-formation (Hutchings, Crampton \& Johnson 1995). However, to determine whether LQGs trace large-scale structure require observations that can identify associated clustering of quiescent galaxies, not just star-forming regions, as these are likely to be short-lived and do not necessarily imply significant associated masses.

 The quiescent galaxies which are the easiest to detect, and which are the best indicators of galaxy clustering, are the massive ellipticals which dominate cluster cores. These have been observed at low redshifts to form a homogeneous population, forming very tight colour-magnitude relations known as {\em red sequences} indicative of old stellar populations (12--13 Gyr). 
 This red sequence has been followed in the optical for clusters out to $z\simeq0.9$ and is consistent with the passive evolution of galaxies that formed in a monolithic collapse at $z\ga3$. At $z\ga1$ these galaxies should be characterised by extremely-red optical-NIR colours ($I-K\simeq4$) as their strong 4000\AA$\,$ break is redshifted into the $I$ band. Several $z\simeq1.2$ clusters have been found through the combination of ultra-deep optical and deep near-infrared imaging of targetted fields around high-redshift AGN (Dickinson 1995; Tanaka et al. 2000) and regions of extended X-ray emission (Stanford et al. 1997; Rosati et al. 1999), with red sequences observed at $I-K\simeq4,R-K\simeq6$ as predicted by the monolithic collapse model.

To examine the galaxy environment of quasars in the Clowes-Campusano LQG, we have conducted an ultra-deep optical (to $V\sim27,I\sim26$) study of a $30'\times30'$ field containing 3 quasars from this LQG, using the BTC camera on the 4-m Blanco telescope at CTIO. We have also obtained $K$ imaging of subfields around 2 of the LQG quasars, and have examined the galaxy environments of the quasars by searching for galaxies with the extremely-red optical-NIR colours ($I-K\simeq4,V-K\simeq7$) of passively-evolving galaxies at the quasar redshift.

\begin{figure}[t]
\plotone{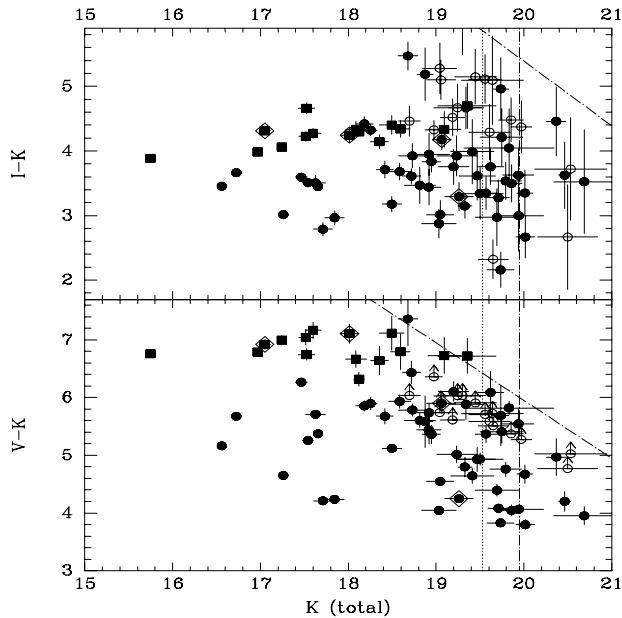}
\caption{Colour-magnitude diagrams for the z=1.226 LQG quasar field. Solid (empty) symbols represent those galaxies detected in $I, K$ and (but not) $V$. The galaxies which make up the red sequences apparent at $V-K\simeq6.9$ and $I-K\simeq4.3$ have square symbols.}
\end{figure}

\section{Galaxy Clustering around a z=1.226 LQG Quasar}

The first quasar environment we have studied in detail is that of the radio-quiet LQG quasar 104420.8+055739 at z=1.226 for which we obtained UKIRT imaging of a $2.25'\times2.25'$ field centred on the quasar, reaching $K=20$. We find a  3.5$\sigma$ excess of $K<20$ galaxies in this field, with 79 observed against 41 expected. This excess is due entirely to a factor $\sim$11 overdensity of $I-K>3.75$ galaxies, that must have $z\ga0.8$ to explain their colour. Half of the $I-K>3.75$ galaxies make up the clear red sequences in the colour-magnitude plots of Figure 1 at  $V-K\simeq6.9, I-K\simeq4.3$, comparable to red sequences observed in other $z\simeq1.2$ clusters (Dickinson 1995; Rosati et al. 1999). This indicates a population of 15--18 passively-evolving massive ellipticals at the quasar redshift, and is clear evidence of a moderately rich cluster associated with the quasar. The remainder of the $I-K>3.75$ galaxies, mostly with $K\ga19$, have much bluer optical colours than the red sequence galaxies with $V-I<2.00$. These do not appear to fit any of the standard Bruzual \& Charlot (1993) GISSEL96 evolutionary models which assume: either a single burst of star-formation followed by passive evolution; or an exponentially-decaying star-formation rate. Such galaxies appear rare in field regions ($\la 0.5\,{\rm arcmin}^{-2}$), but have been observed both in other $z\simeq1.2$ clusters (Tanaka et al. 2000) and at fainter magnitudes with $20\la K\la22$ (Moustakas et al. 1997), and have been described as `red outliers'. Given their relative rarity in field regions, we assume that they are associated with the cluster, and that they have some recent star-formation that causes the blue optical colour, and that their extremely-red $I-K$ colour is produced by a combination of dust and dominant old stellar populations.

\begin{figure}[t]
\plotone{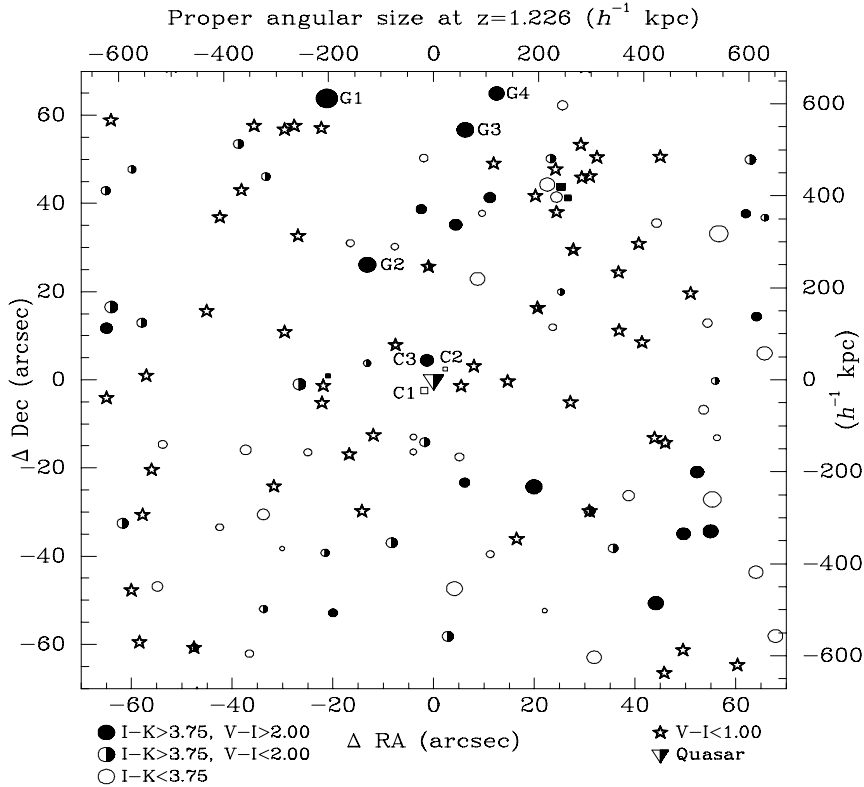}
\caption{Spatial distribution of galaxies in the field of quasar 104420.8+055739.  Solid symbols represent the red sequence galaxies, half-filled symbols represent the red outlier galaxies, and the star symbols represent the blue star-forming galaxies.}
\end{figure}

Having found associated clustering for the z=1.226 LQG quasar, we consider the spatial distribution of the galaxies (Figure 2) to examine the quasar's location in relation to the clustering. The red sequence galaxies (shown as solid symbols) do not appear concentrated towards the quasar. Instead they appear concentrated in two compact groups, one towards the top-centre of the image (including the four brightest members labelled G1--4) and the other in the bottom-right corner. This suggests that the galaxy excesses and red sequences could be due to two clusters at similar redshifts.

There are three galaxies (labelled C1--C3) within 5 arcsec of the quasar. C3 is one of the red sequence galaxies, whilst C1 and C2, both just 3 arcsec from the quasar, are compact and somewhat fainter and bluer. Compact companions are found for a significant fraction of quasars, and it has been suggested that these are the remnants of the galaxy merging event which triggered the quasar. Certainly, if C1 and C2 are associated with the quasar, they have undergone a recent episode of star-formation, presumably caused by the merger process.

To compare this quasar field with the results of Hutchings et al. we have also examined the distribution of blue ($V-I<1.00, I<25$) galaxies (shown as star symbols). We find a concentration of blue galaxies within 30 arcsec of the quasar, which appears to be extended towards the north-east, forming a `band' that bisects the two groups of red sequence galaxies. This band presumably corresponds to a region of enhanced star-formation, similar to those observed by Hutchings et al. around several $z\simeq1.1$ LQG quasars. We propose that this band of star-formation, and possibly the quasar itself, have been triggered by the merging of the two groups of red sequence galaxies, as the galaxies interact with the colliding ICMs. Cluster merging events have been observed at low redshifts to be capable of inducing starbursts simultaneously in galaxies across cluster length-scales (e.g. Caldwell et al. 1993). Such events are predicted to be relatively common at high redshifts in hierarchical clustering models, and many of the high redshift clusters found to date show signs of merging.

\begin{figure}[t]
\plotone{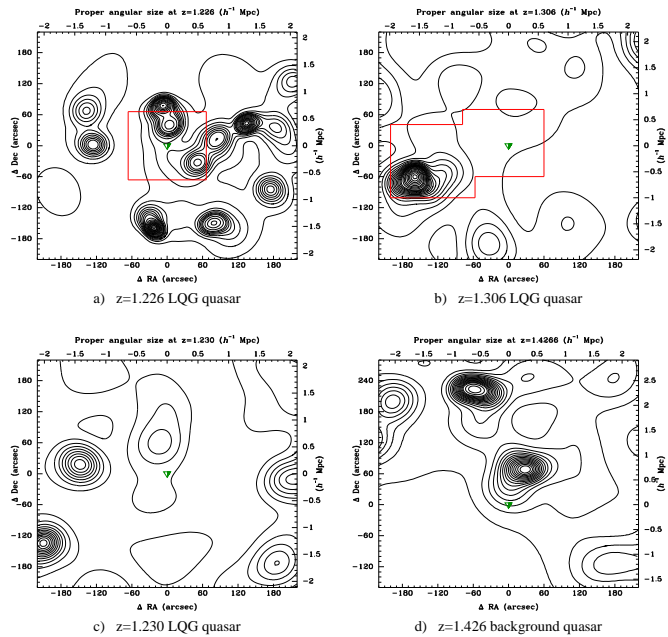}
\caption{Density distribution of red galaxies with $V-K>2.25,$ $I<23.50$ in $7'\times7'$ fields centred on the four $z\simeq1.3$ quasars (3 LQG and one background) in the BTC image. The boxes indicate the regions covered by $K$ imaging.}
\end{figure}

\section{The Galaxy Environment of Four Quasars at z$\simeq$1.3}

To examine the galaxy environments of all four $z\simeq1.3$ quasars in the BTC field, and to search for any large-scale structure associated with the LQG, we have selected {\em red} galaxies with the same optical colours and magnitudes as the red sequence galaxies ($V-I>2.25,I<23.5$) found around the z=1.226 quasar. The selection criteria should pick out $z\ga0.5$ passively-evolving galaxies, and so peaks in their density distribution (as estimated using the adaptive kernel method) should mark out $z\ga0.5$ clusters. This can be considered a crude example of the red sequence cluster finding method successfully used by Gladders \& Yee (2000). Density maps of $7\times7\,{\rm arcmin}^{2}$ ($4\times4\,h^{-2}\,{\rm Mpc}^{2}$ at $z\simeq1.3$) fields around each of the four $z\simeq1.3$ quasars are shown in Figure 3.

The density map of the z=1.226 LQG quasar field indicates that the clusters formed by the two groups of red sequence galaxies extend significantly beyond the $K$ image, the south-eastern group extending 3 arcmin to the north and east. A statistical analysis of the two groups reject the null hypothesis that the galaxy distribution is taken from a single elliptical gaussian probability density function at the 2\% level, i.e. the substructure is real. There are two further compact groups to the south of the quasar, and the structure as a whole is suggestive of being a cluster in the early stages of formation though the progressive coalescence of subclusters.

The most significant cluster of red galaxies across the entire BTC field is located 3 arcmin to the south-west of the z=1.306 LQG quasar (Figure 3b), and so $K$ imaging was obtained to determine whether this was associated with the quasar. However photometric redshift estimates based on the galaxies' $VIK$ colours indicate that the cluster lies at $z\simeq0.9$, and not at the quasar redshift, the quasar itself appearing to reside in a poor environment. The cluster may not be associated with the LQG, but still appears interesting. It is dominated by a very compact core containing a dozen galaxies whose colours are consistent with being passively-evolving galaxies at $z\simeq0.9$ but whose morphologies appear unusually compact, as well a comparable number of faint blue galaxies indicating that there is significant star-formation occurring in the cluster.
 The third LQG quasar also appears to reside in a poor environment, with no sign of clustering of optically-red galaxies nearby. A background quasar at z=1.426 however is located on the edge of a cluster of red galaxies, with a second cluster 3 arcmin to the north, suggesting that this quasar is associated with a rich environment.

The four quasars are located in a variety of environments, but those with associated clustering are found on the cluster {\em peripheries} rather than in the cluster cores. This is in agreement with observations of 7 radio-loud quasars at $1\leq z<1.6$ (Sanch\'{e}z \& Gonz\'{a}lez-Serrano 1999), and with the framework of quasar activity being triggered by the infall of gas onto a seed black hole. Firstly, galaxies in the centres of clusters have previously lost most or all of their gas, and secondly, the encounter velocities of galaxies in the cluster cores are too high for mergers to efficiently trigger nuclear activity. The finding of associated clustering of passively-evolving galaxies for some of the LQG quasars supports the theory that LQGs trace large-scale structure at high-redshifts, although the relationship between quasars and mass/galaxies appears complex.


\begin{references}
\reference Bruzual, G., \& Charlot, S. 1993, ApJ, 405, 538
\reference Caldwell, N. et al. 1993, AJ, 106, 473
\reference Clowes, R.~G., \& Campusano, L.~E. 1991, MNRAS, 249, 218
\reference Crampton, D., Cowley, A. P., \& Hartwick, F.~D.~A. 1989, ApJ, 345, 59
\reference Dickinson, M. 1995, ASP Conf. Ser. Vol. 86, Fresh View of Elliptical Galaxies, ed. A. Buzzoni, A. Renzini \& A. Serrano (San Fransisco: ASP), 286
\reference Gladders, M.~D. \& Yee, H.~K.~C. 2000, AJ in press, (astro-ph/0004092)
\reference Haines, C.~P., Clowes,~R. G., Campusano, L.~E., \& Adamson, A.~J., 2000, \\ MNRAS in press, (astro-ph/0011415)
\reference Hutchings, J.~B., Crampton, D., \& Johnson, A. 1995, AJ, 109, 73
\reference Moustakas, L.~A. et al. 1997, ApJ, 475, 443
\reference Rosati, P. et al. 1999, AJ, 118, 76
\reference Sanch\'{e}z, S.~F. \& Gonz\'{a}lez-Serrano, J.~I. 1999, A\&A, 352, 383
\reference Stanford, S.~A. et al. 1997, AJ, 114, 2232
\reference Tanaka, I. et al. 2000, ApJ, 528, 123
\reference Webster, A. 1982, MNRAS, 199, 683 
\end{references}
\end{document}